\documentclass[12pt,preprint]{aastex}

\def\kms{\ifmmode{\rm km\thinspace s^{-1}}\else km\thinspace s$^{-1}$\fi}
\def\ms{\ifmmode{\rm m\thinspace s^{-1}}\else m\thinspace s$^{-1}$\fi}

\shortauthors{Konacki et al.}
\shorttitle{OGLE-TR-10}

\begin{document}

\title{A transiting extrasolar giant planet around the star OGLE-TR-10}

\author{Maciej Konacki\altaffilmark{1}, Guillermo Torres\altaffilmark{2},
Dimitar D.\ Sasselov\altaffilmark{2} and Saurabh Jha\altaffilmark{3}}

\altaffiltext{1}{California Institute of Technology, Div.\ of
Geological \& Planetary Sciences 150-21, Pasadena, CA 91125, USA;
e-mail: maciej@gps.caltech.edu}

\altaffiltext{2}{Harvard-Smithsonian Center for Astrophysics, 60
Garden St., Cambridge, MA 02138, USA}

\altaffiltext{3}{Department of Astronomy, 601 Campbell Hall, 
University of California, Berkeley, CA 94720, USA}

\begin{abstract}

We report a transiting extrasolar giant planet around the star
OGLE-TR-10 (orbital period = 3.1 days), which was uncovered as a
candidate by the OGLE team in their photometric survey towards the
Galactic center \citep{Udalski:02a}. We observed OGLE-TR-10
spectroscopically over a period of two years (2002--2004) using the
HIRES instrument with an iodine cell on the Keck~I telescope, and
measured small radial velocity variations that are consistent with the
presence of a planetary companion. This confirms the earlier
identification of OGLE-TR-10b by our team and also recently by
\cite{Bouchy:04b} as a possible planet.  Additionally, in this paper
we are able to rule out a blend scenario as an alternative
explanation. From an analysis combining all available radial velocity
measurements with the OGLE light curve we find that OGLE-TR-10b has a
mass of $0.57 \pm 0.12$~M$_{\rm Jup}$ and a radius of $1.24 \pm
0.09$~R$_{\rm Jup}$.  These parameters bear close resemblance to those
of the first known transiting extrasolar planet, HD~209458b. 

\end{abstract}


\section{Introduction}

Photometric observations combined with radial velocity measurements of
stars harboring transiting giant planets yield the absolute radii,
masses and densities of the planets.  The ongoing photometric surveys
for transiting planets have now delivered large numbers of candidates,
but those observations alone cannot distinguish between companions
that are Jupiter-size planets, brown dwarfs, or late type M dwarfs,
since they all have similar radii (of order 0.1 $R_{\sun}$). They also
cannot distinguish between true transiting planets and so-called
``blends" --- chance alignments or physical triple systems involving
an eclipsing binary that can mimic planetary transits
\citep{Konacki:03a,Torres:04b,Torres:05}. These can be very common,
particularly in crowded fields. Hence follow-up spectroscopic
observations are not only necessary to measure the radial velocities
and derive the mass of a transiting planet, but they are also vital
for weeding out astrophysical false positives.

The first confirmed case of a transiting planet proposed by a
photometric survey \citep{Udalski:02a,Udalski:02b} was that of
OGLE-TR-56b \citep{Konacki:03a, Torres:04a, Bouchy:04b}.  Three other examples
have also come out of the OGLE survey \citep{Udalski:02c,Udalski:03}:
OGLE-TR-113b \citep{Bouchy:04a,Konacki:04}, OGLE-TR-132b
\citep{Bouchy:04a,Moutou:04} and OGLE-TR-111b \citep{Pont:04}.
Additionally, the first transiting planet from a wide-field
bright-star survey has recently been reported
\citep[TrES-1;][]{Alonso:04}. This brings the total number to six,
including the case of HD~209458b, a planet originally discovered in
the course of a radial velocity survey and only later found to undergo
transits \citep{Henry:00, Charbonneau:00}. 

OGLE-TR-10 was identified as a promising candidate by the OGLE team
during their 2001 campaign in three fields towards the Galactic center
\citep{Udalski:02a}.  The possible planetary nature of its companion
based on spectroscopic follow-up was first established by
\cite{Konacki:03b}. In that paper we reported a tentative radial
velocity semi-amplitude of $K = 100\pm43$ \ms, and a mass for the
putative planet of $M_p = 0.7 \pm 0.3$ $M_{\rm Jup}$. However, the
possibility of a blend could not be categorically ruled out at the
time due to the small number of observations.  Recently
\cite{Bouchy:04b} also called OGLE-TR-10 a possible planet, with $K =
81 \pm 25$ \ms\ and $M_p = 0.66 \pm 0.21$ $M_{\rm Jup}$, although
these authors were still unable to completely exclude a blend scenario
because of the insufficient signal-to-noise ratio (SNR) of their
observations. 

In this paper we present the results from additional spectroscopic
monitoring of OGLE-TR-10 with Keck~I/HIRES on two additional runs, for
a total of three seasons (2002--2004). We are now able to confirm the
planetary nature of OGLE-TR-10b. Our new radial velocities and the
parameters of the parent star are presented in
Section~\ref{sec:observations} and Section~\ref{sec:spectrum}. The
OGLE light curve solution and planet parameters are derived in
Section~\ref{sec:orbit}, where we combine our radial velocities with
those reported by \cite{Bouchy:04b}. In Section~\ref{sec:falsepositive} we
analyze and rule out a blend as an alternative explanation. The
results are discussed in Section~\ref{sec:discussion}. 

\section{Observations}
\label{sec:observations}

OGLE-TR-10 was observed spectroscopically with the Keck~I telescope
using the HIRES instrument \citep{Vogt:94::} in July 2002, August
2003, and July 2004.  The exposure times ranged from 30 to 50 minutes
and the wavelength coverage was 3850\,\AA\ to 6200\,\AA\ (36 echelle
orders), at a resolving power of $R \simeq 65,\!000$. We collected 10
spectra of which 9 were taken with the iodine gas absorption cell
(I$_2$) that superimposes a dense forest of absorption lines directly
on the stellar spectrum in the region from approximately 5000\,\AA\ to
6200\,\AA\ (some 12 echelle orders). The iodine orders had typical SNR
of 15 to 25 per pixel.  One spectrum was taken without
the I$_2$ to serve as the template for the iodine velocity reductions.

For faint stars such as OGLE-TR-10 ($V = 15.8$, $I = 14.9$) light
losses due to the iodine cell typically prevent one from using that
technique to determine precise radial velocities in the manner done
for brighter Doppler targets. However, as described by
\cite{Konacki:03b} precision as good as 50~\ms\ can still be achieved
with the I$_2$ cell on faint targets by using a synthetic spectrum
instead of an observed spectrum as the template. Additionally, in
order to monitor changes in the instrumental profile (PSF) that affect
the velocities significantly, we obtained observations a bright star
with the I$_2$ cell on every night we observed OGLE-TR-10. Those
observations were used to model the PSF and establish the parameters that
we then applied to OGLE-TR-10. In this way we were able to determine
radial velocities with typical uncertainties of about 60~\ms, which
are listed in Table~\ref{tab:rvs}. Our spectroscopic orbital solution
is described below in \S\ref{sec:orbit}. 

\section{Parameters of the parent star}
\label{sec:spectrum}

The stellar parameters for OGLE-TR-10 were derived from our
high-resolution co-added Keck spectrum (SNR of 44) with fits to
synthetic spectra computed from model atmospheres for different
compositions based on the ATLAS~9 and ATLAS~12 programs by
\cite{Kurucz:95::}. We use a code re-written in Fortran-90 (J.\
Lester, priv.\ comm.) and incorporating new routines for improved
treatment of contributions from various broadening mechanisms, as well
as updated and expanded opacities and line lists.  This code has been
tested extensively and performs very well for solar-type stars (F-K
type) such as OGLE-TR-10. The fits between observed and synthetic
spectra were made in spectral regions unaffected by the I$_2$ lines,
and including a large number of metal absorption lines of different
ionization and excitation states as well as the core and wings of the
$\lambda$4861 H$\beta$ line. Our effective temperature determination,
$T_{\rm eff}=5750$~K, has an estimated accuracy of about 100~K (see
below).  Our projected rotational velocity, $v \sin i = 3$~\kms, is
good to better than 2~\kms; we use an approach to the line broadening
similar to that described by \cite{FV:03::}. Our procedures allow us
to derive the metallicity, [Fe/H], and the limb darkening parameter,
$u_I = 0.51 \pm 0.04$, with good confidence. [Fe/H] is estimated to be
solar with a 0.2~dex uncertainty.  The surface gravity is much more
difficult to determine reliably. We estimate $\log g =
4.4^{+0.4}_{-0.9}$, in which the upper limit is much better
constrained than the lower bound, as is common in this type of
analysis. Given the morphology of evolutionary tracks for main
sequence stars and the lack of an independent distance estimate for
OGLE-TR-10, this uncertainty in $\log g$ allows us to rule out a giant
or subgiant status for the star, but provides only a weak constraint
on the stellar radius on the main sequence.

\begin{figure}
\figurenum{1}
\includegraphics[scale=1.00]{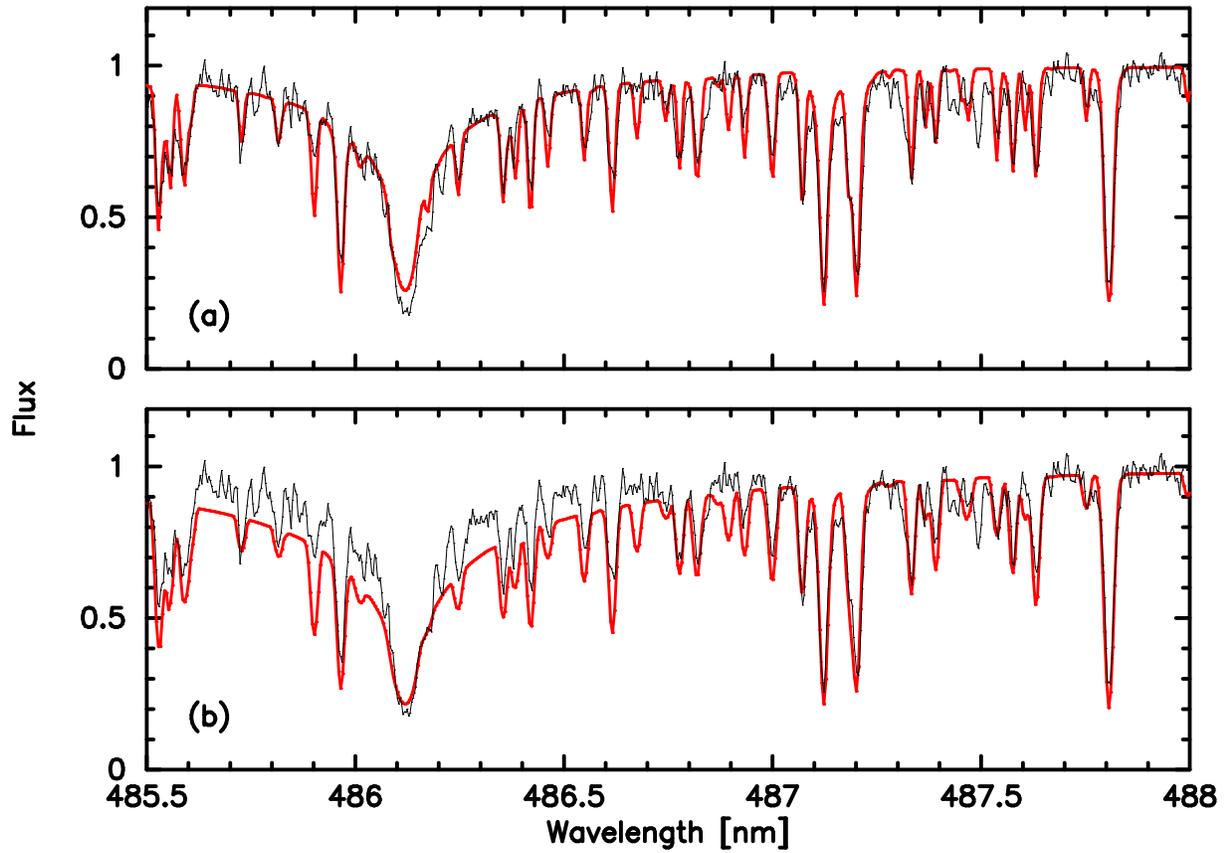}
\caption{Portion of the observed (co-added) spectrum of OGLE-TR-10
around the H$\beta$ line with {\it (a)} our best-fit synthetic
spectrum superimposed (smooth solid line) and {\it (b)} a synthetic
spectrum for the stellar parameters derived by \cite{Bouchy:04b}.
\label{fig:spectrum}}
\end{figure}

The stellar mass and radius were determined using a stellar evolution
code described in detail elsewhere \citep{Cody:02::,S:03::}.  As
evolutionary tracks are nearly vertical in the $T_{\rm eff}$ vs.\
$\log g$ plane for the range of interest for OGLE-TR-10, the largest
uncertainty in deriving the radius would be an error in $T_{\rm eff}$.
A temperature that is too hot would lead to an overestimate of the
stellar mass and also a stellar radius that is too large (given the
stricter upper bound on $\log g$). This in turn would lead to a
significant overestimate of the derived planet radius ($R_p$) in the
solution of the transit light curve.  Figure~\ref{fig:spectrum}
illustrates our efforts to obtain an accurate value of $T_{\rm eff}$,
and in particular the strong sensitivity of Balmer line profiles to
the temperature. We point out that LTE radiative equilibrium codes
like ATLAS are particularly well suited to fit the wings of such
lines. The cores are usually affected by chromospheres and non-LTE
effects, and are expected to appear deeper in a solar-type star like
OGLE-TR-10. A fit to the wings of the $\lambda$4861 H$\beta$ line requires
a careful setting of the contunuum (for better illustration only a
fraction of the HIRES order is shown in Fig.~ref{fig:spectrum}). 
To guard against problems with contunuum setting and line broadening,
in deriving $T_{\rm eff}$ we also rely on temperature-sensitive line
pairs: moderate-strength metal lines having different (often opposite)
sensitivity to temperature and co-located (within 0.1$-$0.2~nm) in
wavelength (see Gray \& Johanson 1991 for the general idea and some
line pairs).  We get $T_{\rm eff}=5800~\pm 100$~K,
in good agreement with the Balmer line wings and neutral-to-ionized metal
lines comparisons. 

Thus, based on the overall fit of the good-quality Keck
spectrum, OGLE-TR-10 is confidently identified as a close analog to
our Sun. Our estimates of the stellar parameters are significantly
different from those reported by \cite{Bouchy:04b} ($T_{\rm eff} =
6220 \pm 140$ K, $\log g = 4.70 \pm 0.34$, [Fe/H] = $+0.39 \pm 0.14$,
$M = 1.22 \pm 0.045$ M$_{\sun}$, $R = 1.21 \pm 0.066$ R$_{\sun}$).
However, as can be also seen in Figure~\ref{fig:spectrum}, our
determinations lead to a much better match to the overall high SNR spectrum of
OGLE-TR-10 from Keck~I/HIRES. 

\section{Analysis and results}
\label{sec:orbit}

\begin{figure}
\figurenum{2}
\hskip 2.5cm \includegraphics[scale=0.9]{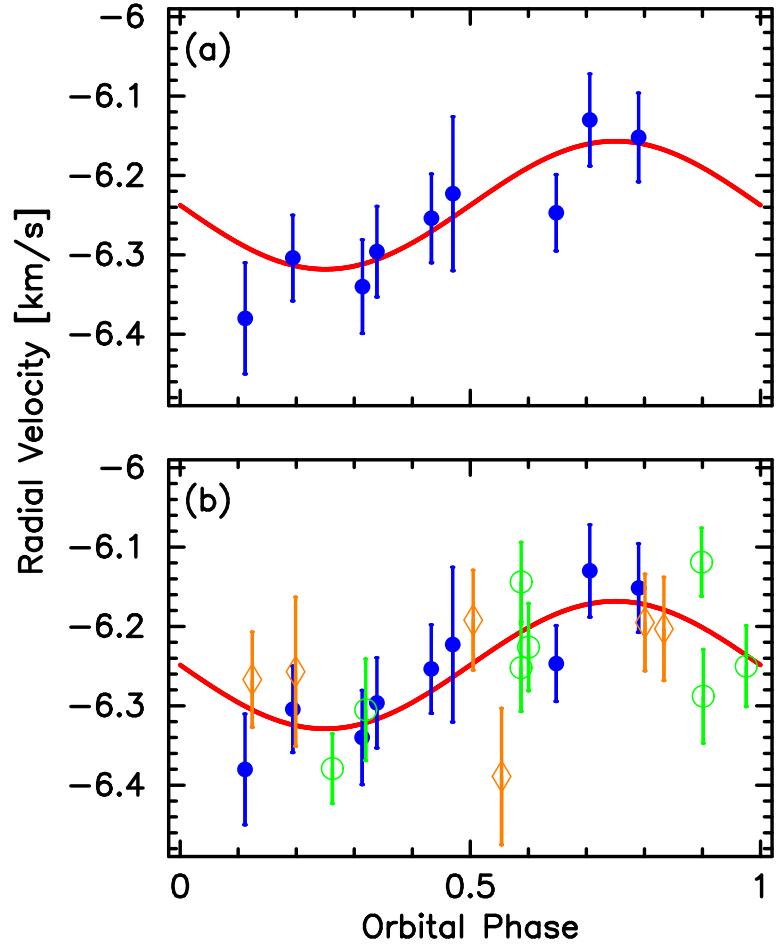}
\caption{Radial velocity measurements and fitted velocity curve for
OGLE-TR-10, as a function of orbital phase. The transit ephemeris is
adopted from the photometry (see text)\label{fig:oglervs}. Our HIRES
velocities are represented with filled circles, and UVES and FLAMES
velocities by \cite{Bouchy:04b} are shown with open diamonds and open
circles, respectively. Panel {\it (a)} is for the fit to HIRES
velocities only (rms of 45~\ms) and panel {\it (b)} is for the fit to
all available data (rms = 60~\ms).\label{fig:rv}}
\end{figure}

Our radial velocities from \S\ref{sec:observations} show clear
variations as a function of photometric phase.  Adopting the orbital
period and epoch as reported for OGLE-TR-10 by \cite{Udalski:02a} we
fitted for a circular orbit solving for the center-of-mass velocity
and the semi-amplitude $K$ (see Figure~\ref{fig:rv}a). The best-fit
value of $K = 77 \pm 23$ \ms\ is consistent with the early result from
\cite{Konacki:03b} and \cite{Bouchy:04b}.
The rms of our spectroscopic solution is
45~\ms, and from Monte Carlo simulations we find that the probability
of obtaining by chance a $K$ amplitude as large as we measure is only
$\sim4 \times 10^{-4}$. 

\begin{figure}
\figurenum{3}
\includegraphics[scale=1.00]{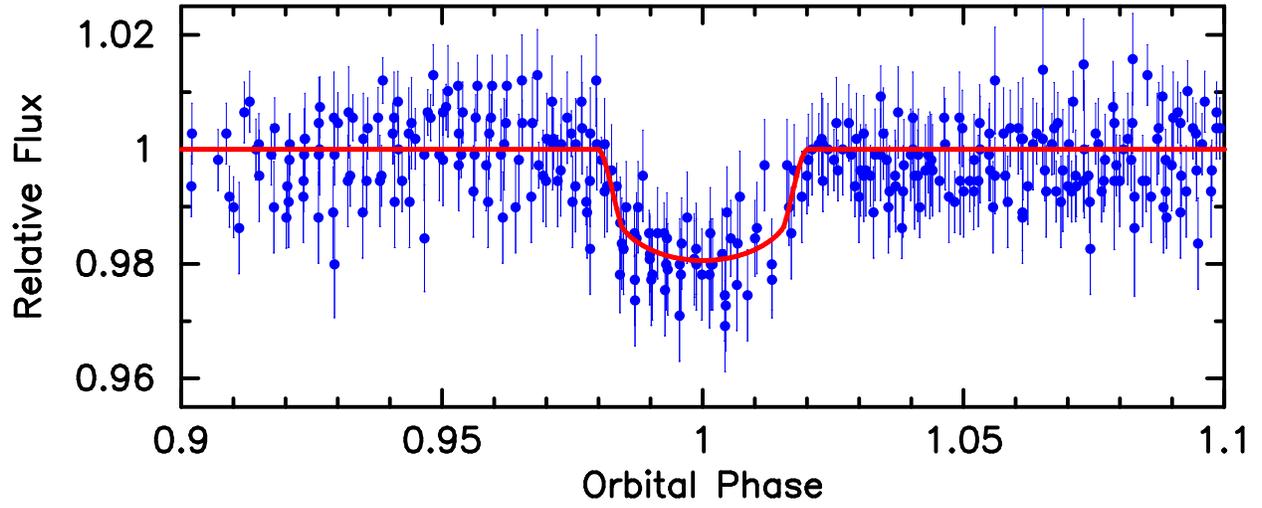}
\caption{OGLE photometry for OGLE-TR-10 in the $I$ band, with our
best fit transit light curve. The resulting parameters are listed in
Table~\ref{tab:results}.\label{fig:lcurve}}
\end{figure}

For our final spectroscopic orbital solution we combined our HIRES
measurements with 14 velocity measurements by
\cite{Bouchy:04b}\footnote{Two additional measurements by these
authors were considered by them to be of lower quality and rejected.
We have done the same here.}.  The latter were obtained with two
different configurations of the VLT spectrograph UVES: in the standard
slit mode (hereafter UVES), and a setup with a fiber link (FLAMES). We
considered these as independent data sets, and solved for the
corresponding velocity offsets relative to HIRES along with the rest
of the orbital elements.  The combined fit gives an improved velocity
semi-amplitude of $K = 80 \pm 17$ \ms, with an rms residual of 60 \ms\
(see Figure~\ref{fig:rv}b).  The corresponding false alarm probability
from Monte Carlo simulations is $\sim10^{-6}$.  We use this combined
fit in the rest of our paper. The best-fit parameters are presented in
Table~\ref{tab:results}, along with our light curve solution based on
the OGLE photometry, using the formalism of \cite{Mandel:02}
(Figure~\ref{fig:lcurve}). By combining this information with the
stellar parameters described previously we derive for the planet in
orbit around OGLE-TR-10 an absolute mass of $0.57 \pm 0.12$~M$_{\rm
Jup}$ and a radius of $1.24 \pm 0.09$~R$_{\rm Jup}$.  The formal
errors include the contribution from uncertainties in the mass and
radius of the parent star. 

\section{False positive rejection}
\label{sec:falsepositive}

To test for the possibility that the radial velocities we measured for
OGLE-TR-10 are due simply to blending with an eclipsing binary, we
computed the spectral line bisectors from our Keck observations as
described by \cite{Torres:04a}. There is no significant variation of
the bisector spans as a function of phase (see below), as would be
expected if lines from another star were causing asymmetries in the
profiles of the main star by moving back and forth with the
photometric period.  Similar results were reported by
\cite{Bouchy:04b}. 

\begin{figure} 
\figurenum{4} 
\hskip 2.0cm \includegraphics[scale=0.6]{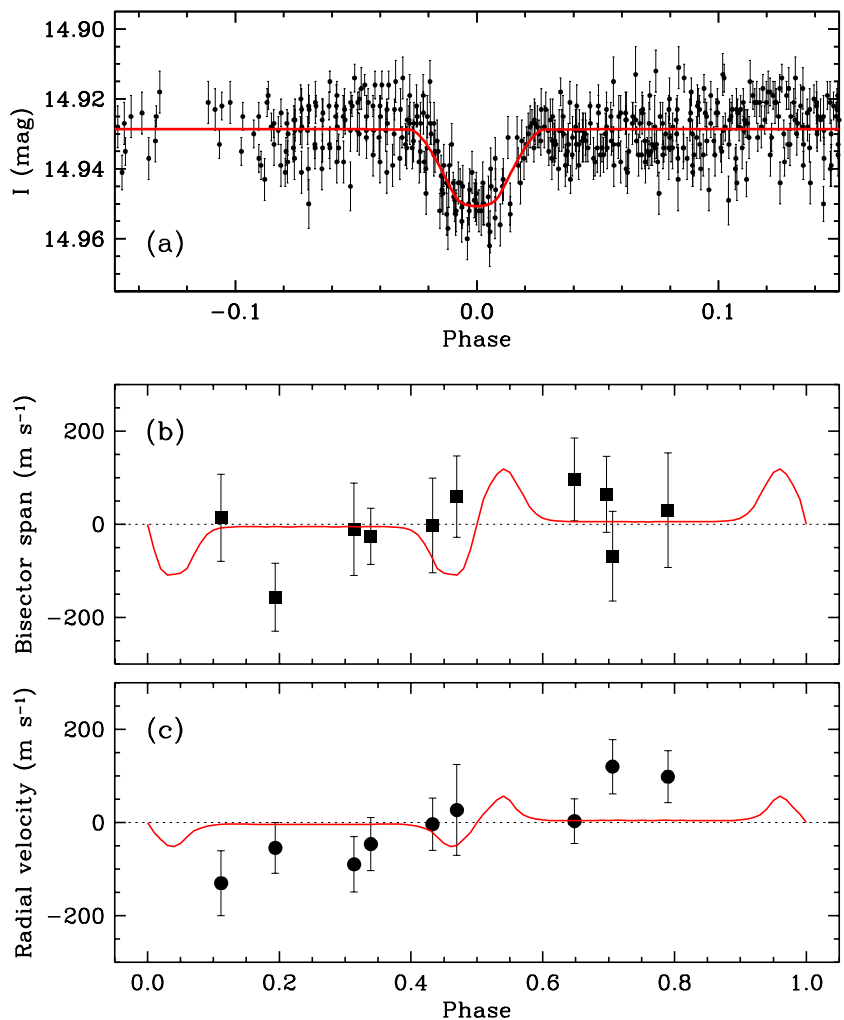}
\vskip 2cm 
\caption{Blend model for OGLE-TR-10 in which an eclipsing binary (F9V
+ K5V) is located in the background of the main G star, and has a
relative brightness (4.5\%) that makes it undetectable in our spectra.
This particular model assumes a systemic velocity for the binary equal
to the velocity of the G star, but extensive tests show that this has
no effect on the conclusions. {\it (a)} Fit to the OGLE light curve
near the primary eclipse. {\it (b)} Measured bisector spans as a
function of phase, compared to predictions from the blend model. The
cusps are the result of the large velocity semi-amplitude of the
eclipsing binary compared to the width of the lines of the G star
\citep[see][]{Torres:05}. {\it (c)} Our radial velocity measurements
shown with the predictions from the blend model. The observed velocity
trend is not reproduced by the model. \label{fig:figblend}}
\end{figure}

Next we modeled the OGLE light curve in detail using the procedures
described \cite{Torres:04b}, assuming a configuration consisting of an
eclipsing binary blended with the main G star forming a hierarchical
triple system.  Although we were able to achieve an excellent fit that
is essentially indistinguishable from a true transit light curve, this
blend model predicts an optical brightness for the primary of the
eclipsing binary that is \emph{greater} than the G star itself, which
is clearly not observed.  We then relaxed the condition that the three
stars be at the same distance, and considered models with the
eclipsing binary in the background in order to make it fainter. Here
too we were able to find a perfectly acceptable fit for an eclipsing
binary composed of an F9V star and a K5V star located several kpc
behind the G star (see Figure~\ref{fig:figblend}a).  The relative
brightness of the eclipsing pair in this model is only about 4.5\%
compared to the G star, which would be just below our threshold of 5\%
for detecting lines of another star in the spectra
\citep[see][]{Konacki:03b}. The predicted orbital velocity
semi-amplitude of the F9 star is 65~\kms, and its spectral lines
should show a rotational broadening corresponding to $v \sin i =
20$~\kms\ (assuming the spin is synchronized with the orbital motion). 

In order to test this blend scenario further we ran extensive
numerical simulations following \cite{Torres:05} to predict the
bisector span and radial velocity variations that would be expected
from this configuration. In Figure~\ref{fig:figblend} we compare these
predictions with the observations as a function of orbital phase. The
expected bisector span variations in Figure~\ref{fig:figblend}b are
relatively small (less than about $100$~\ms), and are therefore still
consistent with the measurements, which show no significant variation
given typical errors that are also about $100$~\ms.  Thus, the fact
that bisector spans for a transit candidate show no appreciable change
with phase \emph{does not} necessarily rule out a blend
scenario. However, Figure~\ref{fig:figblend}c indicates that the
expected radial velocity variations are even smaller, and do not show
the trend with phase displayed by the velocity measurements for
OGLE-TR-10. As discussed in \S\ref{sec:orbit} the latter velocity
trend is confirmed and reinforced by independent measurements from two
different data sets, as reported by \cite{Bouchy:04b}. The false alarm
probability based on Monte Carlo simulations that a set of velocity
measurements resulting from such a blend scenario would have the phase
coherence we observe, with a semi-amplitude as large as we observe, is
$1.3\times 10^{-3}$ from our HIRES measurements alone, and $3 \times
10^{-5}$ using all the RV data. This effectively rules out this blend
configuration.  Experiments show that it is not possible to make the
predictions more consistent with all of the observational constraints
by changing the parameters of the blend model. Additional evidence
against a blend has been obtained recently by \cite{Holman:05} based
on new high-quality and high-cadence photometric observations of the
star.  Their accurate light curves in two passbands define the
morphology of ingress and egress significantly better than the OGLE
measurements, and are clearly inconsistent with the light curve from a
blend configuration such as we described above, involving a background
eclipsing binary. We conclude that the observations do not support a
blend scenario, and the planetary nature of the companion to
OGLE-TR-10 is confirmed.

\section{Discussion}
\label{sec:discussion}

\begin{figure}
\figurenum{5}
\includegraphics[scale=0.8]{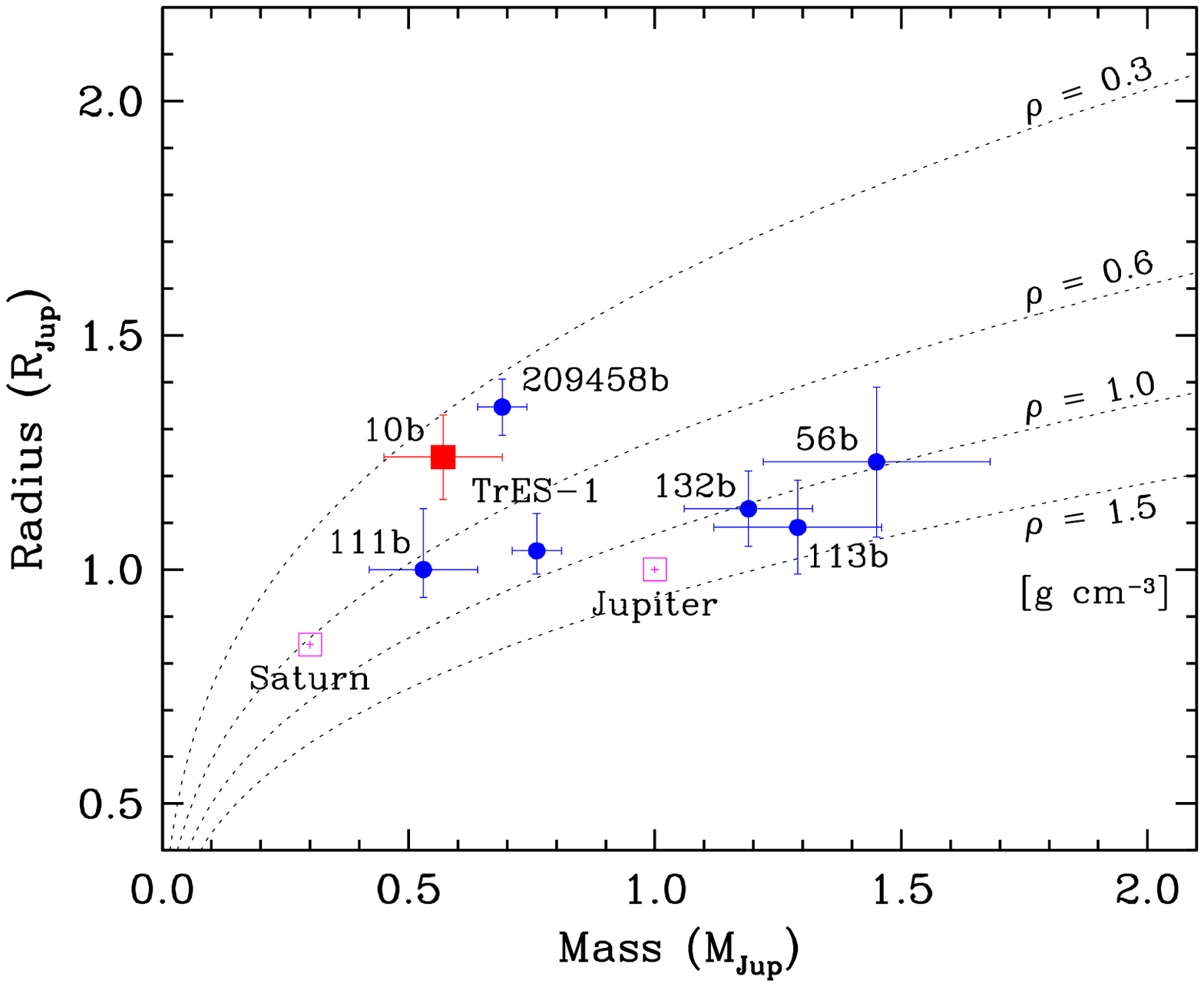}
\vskip -1.5cm
\caption{Radii of transiting extrasolar planets plotted against their
masses. Jupiter and Saturn are included for reference, along with
dotted lines of constant density. Data are from \cite{Brown:01} for
HD~209458b, \cite{Torres:04a} for OGLE-TR-56b, \cite{Moutou:04} for
OGLE-TR-132b, \cite{Pont:04} for OGLE-TR-111b, \cite{Sozzetti:04} for
TrES-1, and this paper for OGLE-TR-10b. For OGLE-TR-113b we have
combined the original observations reported by \cite{Bouchy:04a} and
\cite{Konacki:04}, and derived an improved planetary mass of $1.29 \pm
0.17$ M$_{\rm Jup}$ and a radius of $1.09 \pm 0.10$ R$_{\rm Jup}$ (see
{\tt http://www.gps.caltech.edu/$\sim$maciej/Planets/OGLE-TR-113b.html}).
\label{fig:massrad}}
\end{figure}

OGLE-TR-10b is the seventh known extrasolar transiting planet and the
fifth to come out of the OGLE survey. Admittedly the faintness of the
transiting planet candidates from the OGLE survey requires some of the
largest available telescopes merely to confirm their planetary status,
and other detailed follow-up studies are very difficult to pursue with
current instrumentation.  Nevertheless, these discoveries around faint
stars have been extremely important in the field of extrasolar
planets.  They account for most of the points in the current
mass-radius diagram for giant planets (see Figure~\ref{fig:massrad}),
which relates key properties of these objects for our
theoretical understanding of their structure and evolution.
Additionally, they have led to the discovery of a new class of ``very
hot Jupiters" \citep{Bouchy:04a,Konacki:03a,Konacki:04} with remarkably short
orbital periods (1--2 days).  The importance of faint transit
candidates from surveys like OGLE is likely to continue in the near
future, especially given that the OGLE team has recently released a
new set of 40 candidates that seem very promising \citep{Udalski:04}.
The wide-field surveys, on the other hand, have only recently produced
their first transiting planet \citep{Alonso:04} despite having been in
operation for longer than the OGLE effort. They would provide planets
for more detailed follow-up studies in the future.

OGLE-TR-10b is very similar to HD~209458b \citep{Brown:01} in terms
of its orbital and physical parameters. In particular, its orbital
period (3.1 days) places it in the ``hot Jupiter" category (planets
with periods of 3--4 days) already well populated from the radial
velocity surveys.  OGLE-TR-10b has a low average density and like
HD~209458b might require an additional heating mechanism to explain
it. There are now as many hot Jupiters as there are very hot Jupiters
among the transiting planets uncovered photometrically, and so the
apparent lack of longer-period planets in the transit surveys that
initially appeared to cause some concern \citep{Bouchy:04a, Pont:04}
no longer seems to be a problem, as anticipated by \cite{Gaudi:04}. 

We note, finally, an apparent dichotomy in the mass-radius diagram of
Figure~\ref{fig:massrad} in that the 4 planets with the longer periods
(in the hot Jupiter class) all have small masses ($\sim$0.7~M$_{\rm
Jup}$), while all the short-period planets (very hot Jupiters) have
masses roughly twice as large. This trend, noted previously by
\cite{Mazeh:05}, now seems to be reinforced, and may perhaps be
related to issues of survival of planets in very close
proximity to their parent stars. 

\acknowledgements

MK\ acknowledges support by NASA through grant NNG04GM62G and partial
support by the Polish Committee for Scientific Research, Grant
No.~2P03D~001~22.  GT\ acknowledges support for this work from STScI
program GO-9805.02-A, NASA Origins grant NNG04LG89G, and the Keck PI
Data Analysis Fund (JPL 1257943). SJ\ thanks the Miller Institute for
Basic Research in Science at UC Berkeley for support through a
research fellowship. Support from the Harvard Center for Planetary
Astrophysics is acknowledged. The data presented herein were obtained at the
W.\ M.\ Keck Observatory, which is operated as a scientific
partnership among the California Institute of Technology, the
University of California and the National Aeronautics and Space
Administration. The Observatory was made possible by the generous
financial support of the W.\ M.\ Keck Foundation. This research has
made use of NASA's Astrophysics Data System Abstract Service.

\clearpage

\clearpage

%
%
 
\begin{deluxetable}{cccc}
\tablenum{1}
\tablewidth{21pc}
\tablecaption{Radial velocity measurements for OGLE-TR-10, in the
barycentric frame.\label{tab:rvs}}
\tablehead{
\colhead{HJD} & \colhead{} & \colhead{Velocity} & \colhead{Error} \\
\colhead{(2,400,000+)} & \colhead{Phase} & \colhead{($\kms$)} & \colhead{($\kms$)} }
\startdata
52481.7946 & 0.706 & $-$6.130 &  0.058 \\
52483.7578 & 0.330 & $-$6.296 &  0.057 \\
52853.7813 & 0.648 & $-$6.247 &  0.048 \\
52855.8454 & 0.314 & $-$6.340 &  0.059 \\
52864.7786 & 0.194 & $-$6.304 &  0.054 \\
53206.7856 & 0.470 & $-$6.223 &  0.097 \\
53207.7794 & 0.790 & $-$6.152 &  0.056 \\
53208.7772 & 0.112 & $-$6.380 &  0.070 \\
53209.7727 & 0.433 & $-$6.254 &  0.056 \\
\enddata
\end{deluxetable}

%
%

\begin{deluxetable}{lc}
\tablenum{2}
\tablewidth{36.85pc}
\tablecaption{Orbital and physical parameters for OGLE-TR-10 and its planet.
\label{tab:results}}
\tablehead{
\colhead{\hfil ~~~~~~~~~~~~~~~~~~~~~Parameter~~~~~~~~~~~~~~~~~~~~~~} &  \colhead{Value} }
\startdata
\vspace{2pt}
~~Orbital period (days)\dotfill        & 3.101386~$\pm$~0.000030   \\
~~Transit epoch (HJD$-$2,400,000)\dotfill & 52070.2223~$\pm$~0.0028\phm{2222}       \\
~~Center-of-mass velocity (km~s$^{-1}$)\dotfill    & $-$6.250~$\pm$~0.020\phm{$-$} \\
~~Eccentricity (fixed)\dotfill                 &  0              \\
~~Velocity semi-amplitude (\ms)\dotfill        &  80~$\pm$~17              \\
~~Velocity offset between HIRES and FLAMES (\ms)\dotfill  & $-$14~$\pm$~27\phm{$-$}              \\
~~Velocity offset between HIRES and UVES (\ms)\dotfill  & $+$218~$\pm$~34\phm{$+$2}              \\
\vspace{10pt}
~~Inclination angle (deg)\dotfill        &  89.2~$\pm$~2.0\phn  \\
~~Stellar mass (M$_{\sun}$) (adopted) \dotfill  &  1.00~$\pm$~0.05 \\
~~Stellar radius (R$_{\sun}$) (adopted) \dotfill  &  1.00~$\pm$~0.10 \\
~~Fractional radius ($R_{\rm planet}/R_{\rm star}$)  \dotfill  & $0.127\pm0.017$ \\
\vspace{10pt}
~~Limb darkening coefficient ($I$ band)\dotfill  & 0.51~$\pm$~0.04 \\
~~Planet mass (M$_{\rm Jup}$)\dotfill          &  0.57~$\pm$~0.12      \\
~~Planet radius (R$_{\rm Jup}$)\dotfill        &  1.24~$\pm$~0.09      \\
~~Planet density (g~cm$^{-3}$)\dotfill     &   0.38~$\pm$~0.10 \\
~~Semi-major axis (AU)\dotfill                  &  0.04162~$\pm$~0.00069           \\
\enddata
\end{deluxetable}


\newpage

\begin{thebibliography}{}

\bibitem[Alonso et al.(2004)]{Alonso:04} Alonso, R.\ et al.\ 
2004, \apjl, 613, L153

\bibitem[Bouchy et al.(2004a)]{Bouchy:04a} Bouchy, F., Pont, F., 
Santos, N.\ C., Melo, C., Mayor, M., Queloz, D., \& Udry, S.\ 2004a, \aap, 
421, L13

\bibitem[Bouchy et al.(2004b)]{Bouchy:04b} Bouchy, F., Pont, F., 
Melo, C., Santos, N.\ C., Mayor, M., Queloz, D., \& Udry, S.\ 2004b, 
submitted to \aap, astro-ph/0410346

\bibitem[Brown et al.(2001)]{Brown:01} Brown, T.\ M., 
Charbonneau, D., Gilliland, R.\ L., Noyes, R.\ W., \& Burrows, A.\ 2001, 
\apj, 552, 699

\bibitem[Charbonneau et al.(2000)]{Charbonneau:00}
 Charbonneau, D., Brown, T.\ M., Latham, D.\ W., \& Mayor, M. 2000,
\apjl, 529, L45

\bibitem[Cody \& Sasselov(2002)]{Cody:02::} Cody, A.\ M., \& Sasselov,
D.\ D.\ 2002, \apj, 569, 451

\bibitem[Fischer \& Valenti(2003)]{FV:03::} Fischer, D.\ A., \&
Valenti, J. 2003, in Scientific Frontiers in Research on Extrasolar 
Planets, eds.\ D.\ Deming \& S.\ Seager (San Francisco: ASP), ASP
Conf.\ Ser., 294, 117

\bibitem[Gaudi, Seager, \& Mall\'en-Ornelas(2004)]{Gaudi:04} 
Gaudi, B.\ S., Seager, S., \& Mall\'en-Ornelas, G.\ 2004, ArXiv
Astrophysics e-prints, astro-ph/0409443

\bibitem[Henry et al.(2000)]{Henry:00}
 Henry, G.\ W., Marcy, G.\ W., Butler, R.\ P., \& Vogt, S.\ S. 2000,
\apjl, 529, L41

\bibitem[Holman et al.(2005)]{Holman:05} Holman, M.\ et al., 2005, 
in preparation

\bibitem[Konacki et al.(2003a)]{Konacki:03a}
 Konacki, M., Torres, G., Jha, S., \& Sasselov, D.\ D. 2003a, \nat,
421, 507

\bibitem[Konacki et al.(2003b)]{Konacki:03b}
 Konacki, M., Torres, G., Sasselov, D.\ D., \& Jha, S. 2003b, \apj, 
597, 1076

\bibitem[Konacki et al.(2004)]{Konacki:04} Konacki, M.\ et al.\ 
2004, \apjl, 609, L37

\bibitem[Kurucz(1995)]{Kurucz:95::} Kurucz, R.\ L. 1995, ASP
Conf.\ Ser.\ 78: Astrophysical Applications of Powerful New Databases,
205

\bibitem[Mandel \& Agol(2002)]{Mandel:02}
 Mandel, K., \& Agol, E. 2002, \apjl, 580, L171

\bibitem[Mazeh, Zucker, \& Pont(2005)]{Mazeh:05}
 Mazeh, T., Zucker, S., \& Pont, F. 2005, submitted (see
astro-ph/0411701)

\bibitem[Moutou et al.(2004)]{Moutou:04}
 Moutou, C., Pont, F., Bouchy, F., \& Mayor, M.\ 2004, \aap, 424, L31

\bibitem[Naef et al.(2004)]{Naef:04}
 Naef, D., Mayor, M., Beuzit, J.\ L., Perrier, C., Queloz, D., Sivan,
J.\ P., \& Udry, S. 2004, \aap, 414, 351

\bibitem[Pont et al.(2004)]{Pont:04} Pont, F., Bouchy, F., 
Queloz, D., Santos, N.\ C., Melo, C., Mayor, M., \& Udry, S.\ 2004, \aap, 
426, L15

\bibitem[Sasselov(2003)]{S:03::} Sasselov, D.\ D.\ 2003, \apj, 
596, 1327 

\bibitem[Sozzetti et al.(2004)]{Sozzetti:04}
 Sozzetti, A., Yong, D., Torres, G., Charbonneau, D., Latham, D.\ W.,
Allende Prieto, C., Brown, T.\ M., Carney, B.\ W., \& Laird, J.\ B.,
\apj, 616, L167

\bibitem[Torres et al.(2004a)]{Torres:04a}
 Torres, G., Konacki, M., Sasselov, D.\ D., \& Jha, S. 2004a, \apj,
609, 1071

\bibitem[Torres et al.(2004b)]{Torres:04b}
 Torres, G., Konacki, M., Sasselov, D.\ D., \& Jha, S. 2004b, \apj,
614, 979

\bibitem[Torres et al.(2005)]{Torres:05}
 Torres, G., Konacki, M., Sasselov, D.\ D., \& Jha, S. 2005, \apj, in
press (astro-ph/0410157)

\bibitem[Udalski et al.(2002a)]{Udalski:02a}
 Udalski, A., Paczy\'nski, B., \.Zebru\'n, K., Szyma\'nski, M.,
Kubiak, M., Soszy\'nski, I., Szewczyk, O., Wyrzykowski, \L., \&
Pietrzy\'nski, G. 2002a, Acta Astronomica, 52, 1

\bibitem[Udalski et al.(2002b)]{Udalski:02b}
 Udalski, A., \.{Z}ebru\'n, K., Szyma\'nski, M., Kubiak, M.,
Soszy\'nski, I., Szewczyk, O., Wyrzykowski, \L., \& Pietrzy\'nski, G.
2002b, Acta Astronomica, 52, 115

\bibitem[Udalski et al.(2002c)]{Udalski:02c}
 Udalski, A., Szewczyk, O., \.Zebru\'n, K., Pietrzy\'nski, G.,
Szyma\'nski, M., Kubiak, M., Soszy\'nski, I., \& Wyrzykowski, \L.\
2002c, Acta Astronomica, 52, 317

\bibitem[Udalski et al.(2003)]{Udalski:03} Udalski, A., Pietrzy\'nski,
G., Szyma\'nski, M., Kubiak, M., \.Zebru\'n, K., Soszy\'nski, I.,
Szewczyk, O., \& Wyrzykowski, \L.\ 2003, Acta Astronomica, 53, 133

\bibitem[Udalski et al.(2004)]{Udalski:04} Udalski, A., Szyma\'nski,
M.\ K., Kubiak, M., Pietrzy\'nski, G., Soszy\'nski, I., \.Zebru\'n, K.,
Szewczyk, O., \& Wyrzykowski, \L.\ 2004, ArXiv Astrophysics e-prints,
astro-ph/0411543

\bibitem[Vogt et al.(1994)]{Vogt:94::} Vogt, S.\ S.\ et al.\ 1994,
\procspie, 2198, 362

\end{thebibliography}
\end{document}